\documentclass[journal=apchd5,manuscript=article,layout=twocolumn]{achemso}

\usepackage{chemformula} 

\usepackage{amsmath,amssymb,pifont}
\usepackage{color}
\usepackage{graphicx}
\usepackage{pstool}
\usepackage{makecell}
\usepackage{amsfonts} 
\usepackage{dblfloatfix} 
\usepackage{threeparttable}
\usepackage{amsfonts}
\usepackage{multirow}
\usepackage{float}
\usepackage{graphicx,bm}
\usepackage{subfig}
\usepackage{cancel}
\usepackage{pstool,pstricks}

\usepackage[thinlines]{easytable}

\newcommand{\ve}[1]{\mathbf{#1}}
\newcommand{\te}[1]{\overline{\overline{\mathbf{#1}}}}

\author{Karim Achouri}
\email{karim.achouri@epfl.ch}
\author{Ville Tiukuvaara}
\author{Olivier J. F. Martin}

\affiliation[EPFL]
{Nanophotonics and Metrology Laboratory, Institute of Electrical and Microengineering, $\acute{E}$cole Polytechnique F$\acute{{e}}$d$\acute{{e}}$rale de Lausanne, Route Cantonale, 1015 Lausanne, Switzerland.}

\title{Spatial Symmetries in Multipolar Metasurfaces:\\
	From Asymmetric Angular Transmittance to Multipolar Extrinsic Chirality}

\abbreviations{IR,NMR,UV}
\keywords{Metasurface, Symmetries, Nonlocality, Spatial dispersion, Bianisotropy, Reciprocity, Multipoles}

\begin{document}

%
%
%
%
%

\begin{abstract}

We propose a framework that connects the spatial symmetries of a metasurface to its material parameter tensors and its scattering matrix. This provides a simple yet effective way to effortlessly determine properties of a metasurface scattering response, such as chirality or asymmetric transmission, and which of its effective material parameters should be taken into account in the prospect of a homogenization procedure.

In contrast to existing techniques, this approach does not require any a priori knowledge of group theory or complicated numerical simulation schemes, hence making it fast, easy to use and accessible. Its working principle consists in recursively solving symmetry-invariance conditions that apply to dipolar and quadrupolar material parameters, which include nonlocal interactions, as well as the metasurface scattering matrix. The overall process thus only requires listing the spatial symmetries of the metasurface.

Using the proposed framework, we demonstrate the existence of multipolar extrinsic chirality, which is a form of chiral response that is achieved in geometrically achiral structures sensitive to field gradients even at normal incidence.

\end{abstract}


\section{Introduction}

Over the past few years, metasurfaces have gained increasing attention due to their low form factor and incredible light control capabilities. As a consequence, we have seen a plethora of promising applications emerge, such as light refraction~\cite{yu2011}, polarization control~\cite{balthasar2017} and holography~\cite{zheng2015}. Naturally, this has further sparked an increasing interest towards ever more advanced applications, such as optical analog processing using nonlocal interactions~\cite{kwon2018,abdolali2019} or sensing, detection and polarization multiplexing using chirality~\cite{schaferling2012,hentschel2017,raziman2019,kim2021,chen2021}.

These advances have prompted the need for appropriate modeling approaches able to predict the electromagnetic behavior of a metasurface so as to leverage the adequate available degrees of freedom that they offer for an optimal implementation of the desired specifications or simply to achieve higher performance. Conveniently, several metasurface modeling techniques have been proposed throughout the years~\cite{pfeiffer2013,niemi2013,achouri2015,epstein2016b,achouri2021}, which essentially all revolve around the same concept, i.e., replacing the metasurface by a homogeneous sheet of effective material parameters such as impedances, polarizabilities or susceptibilities. These models are all based on a dipolar description of the metasurface electromagnetic response and they typically include bianisotropic responses that are necessary to model chiral effects~\cite{tang2010,caloz2020,caloz2020a}. 

However, despite being generally powerful, these techniques have been shown to be limited when it comes to modeling the angular scattering properties of a metasurface~\cite{achouri2020a,achouri2021b}. This is particularly a problem for applications pertaining to optical analog signal processing where accurate angular scattering control over a large range of incidence angles is crucially important~\cite{kwon2018,abdolali2019}. One of the main reasons for such a limitation is the fact that optical metasurfaces have unit cells that are relatively large compared to the wavelength, especially those based on dielectric resonators, implying that multipolar contributions beyond the dipolar regime start contributing significantly to their scattering response~\cite{achouri2021b}. Since multipolar components, such as dipoles and quadrupoles, have different angular scattering behavior~\cite{graham1990,raab2005,muhlig2011,nanz2016,alaee2019,evlyukhin2019}, it is clear that metasurface modeling techniques solely based on dipolar components cannot properly model the response of a metasurface under an arbitrary illumination, such as an oblique plane wave. Thankfully, extensions to the conventional dipolar metasurface modeling techniques have been recently proposed~\cite{achouri2021b,rahimzadegan2022}. They include higher-order multipolar contributions as well as nonlocal spatially-dispersive responses that are necessary to adequately connect the multipolar components to the fields exciting the metasurface~\cite{tretyakov2001,agranovich2014,simovski2018}.

While multipolar metasurface modeling techniques exhibit promising features in terms of improved modeling accuracy and better physical insight into the scattering process, they also pose a challenging problem. Compared to dipolar modeling, the inclusion of the required nonlocal 3rd and 4th rank effective material tensors introduces a significant number of additional degrees of freedom and makes the practical application of the modeling significantly more difficult and cumbersome. Nevertheless, this problem may be mitigated by considering that many of these degrees of freedom do not play a role in most conventional metasurfaces and can thus be neglected. For instance, parameters that are related to polarization rotation should not be taken into account for metasurfaces that are known to not induce polarization rotation. Knowing which parameters should or should not be included in the model may be decided, in the most simple cases, just by intuitive reasoning, which has been a common practice in the case of dipolar metasurfaces~\cite{achouri2021}. For slightly more complicated cases, for instance those that require bianisotropic responses, it is possible to determine which are the dominant parameters that should be considered by using a series of numerical simulations with a complex scheme of illumination conditions~\cite{varault2013,bernalarango2014,proust2016}. While such approaches are powerful tools to reduce the complexity of the modeling problem, they require a complicated simulation setup and an important computational cost. Finally, it is possible to individually test the parameters within the homogenized model to determine their angular scattering behaviour and check if this matches that which corresponds to the system and therefore may be present~\cite{achouri2021b}. However, this is cumbersome and furthermore still requires intuition of the symmetries of the scattered fields. This therefore begs the need for a simple, fast, rigorous and effective method to determine which multipolar parameters should be considered and which ones should be excluded from the model. 

On the other hand, the use of chiral responses in metasurfaces for sensing, polarization control or asymmetric transmission applications have led to fascinating works investigating the origins of such responses. It was for instance shown that 3D geometrically chiral structures are not necessary to achieve chiral responses~\cite{hecht1994,potts2004,gonokami2005,bai2007,plum2007,arteaga2016,papakostas2003,plum2009,plum2009a,cao2015,okamoto2019,singh2009,menzel2010a,plum2011,li2013a}. This is possible because a chiral response may be obtained either by using 2D chiral scattering particles placed on a substrate~\cite{hecht1994,potts2004,gonokami2005,bai2007,plum2007,arteaga2016}, which breaks the symmetry of the system in the 3rd dimension, or by illuminating asymmetric particles along specific oblique directions: a phenomenon commonly referred to as extrinsic chirality~\cite{papakostas2003,plum2009,plum2009a,cao2015,okamoto2019}. Due to their dependence on the spatial symmetries of the metasurface or the direction of wave propagation of the illumination, these types of exotic effects turn out to be particularly difficult to grasp from intuition alone, which is deemed to be even more arduous in the case of a multipolar metasurface. This therefore suggests the need for a method to effortlessly predict the existence of a chiral response in a given metasurface structure and for a specific illumination condition.

In this work, our goal is to address these two requirements, namely to devise a method to determine the existence of multipolar components and that of chiral responses (or other peculiar scattering effects) in the case of an arbitrary metasurface. To this end, we propose to establish a connection between the spatial symmetries of a metasurface and its effective material parameters as well as its scattering response. 

Clearly, the idea of drawing such a connection is not new, as many works have already proposed to use spatial symmetries to model metamaterials. For instance, concepts pertaining to group theory have been leveraged to associate the point group of a metasurface to its dipolar material parameters~\cite{arnaut1997,dmitriev2000} or to predict the existence of certain responses based on the symmetries of the metamaterial structure~\cite{padilla2007,baena2007,isik2009,yu2019}. Spatial symmetries have also been used in several other contexts, such as the design of photonic crystals~\cite{heonkim2003,dmitriev2005}, their relationships with reciprocity and chirality and bianisotropy~\cite{birss1963,barron1986,dmitriev2004,maslovski2009,dmitriev2013,achouri2021c,dmitriev2021,poleva2022}, the existence of bound-states in the continuum~\cite{overvig2020} or even the implementation of asymmetric nonlinear responses~\cite{kruk2022}. Finally, several works have also shown a connection between the scattering response of a metasurface and its spatial symmetries~\cite{li2000,kahnert2005,dmitriev2011,dmitriev2013a,arteaga2014,kruk2015a,kruk2020,achouri2021a}. These works typically consider the symmetries associated with the superposition of the metasurface structure with the fields that interact with it, leading to the formulation of corresponding Jones, Mueller or scattering matrices.

Based on the existing literature, we aim at providing a simple, straightforward and coherent framework that connects the symmetries, material tensors and scattering response of a metasurface in a way that does not require an extensive knowledge of group theory. Thus, the proposed approach only requires listing the spatial symmetries of a metasurface, which makes it accessible to the largest possible audience. It also does not require performing any complicated numerical simulation, which makes it fast and simple to use. In addition, this framework extends the existing methods by including the presence of quadrupolar and nonlocal responses, which is crucial for properly modeling certain types of chiral responses, as we shall see. Finally, we will not restrict ourselves to the study of chiral responses but instead investigate the complete scattering response of a metasurface by computing its full scattering matrix. In order to facilitate the use of the proposed framework, we also provide a Python script, which is accessible on GitHub~\cite{kagit}, and that simply requires the list of spatial symmetries of the metasurface.

This paper is organized as follows. In Sec.~\ref{sec_dip}, we review the general approach for transforming the dipolar material parameters of a metasurface according to an arbitrary spatial symmetry, which is a crucial step for ultimatily obtaining the material parameters corresponding to the metasurface.  In Sec.~\ref{sec_mult}, we extend that approach to multipolar responses and, in Sec.~\ref{sec_inv}, finally show how to connect spatial symmetries and multipolar material parameters. In Sec.~\ref{sec_Appsym}, we connect together the spatial symmetries of a metasurface and the fields that interact with it to its corresponding scattering matrix. Finally, in Sec.~\ref{sec_examples}, we provide three examples illustrating the application of the proposed framework.

\section{Material tensors and Symmetries}
\label{sec_mat}

The general approach for connecting together the effective material tensors of a metasurface to its spatial symmetries is based on Neumann's principle~\cite{voigt1910,post1978,oliveira2008}. This principle states that if a system, like a crystal or an electromagnetic structure, is invariant under certain symmetry operations, then its physical properties should also be invariant under the same symmetry operations. It follows that the relationships between spatial symmetries and material tensors may be established by deriving symmetry invariance conditions that apply to the multipolar tensors describing the effective electromagnetic response of a metasurface.

We shall next review the conventional method used to derive such invariance conditions in the case of a medium described by dipolar responses. Then, we will extend these conditions to the case of multipolar responses.

It should be noted that throughout this work, we shall consider that a metasurface is an electrically thin array consisting of a subwavelength periodic arrangement of reciprocal scattering particles. The period of the array is considered small enough compared to the wavelength so that no diffraction orders exist (besides the 0th orders in reflection and transmission) irrespectively of the direction of wave propagation and the refractive index of the background media. Under these assumptions, it follows that the electromagnetic response of such a metasurface can be modeled as that of a \emph{homogeneous} and \emph{uniform} sheet of effective material parameters~\cite{tretyakov2003,achouri2021}.

\subsection{In the case of dipolar responses}
\label{sec_dip}

Let us consider a uniform and homogeneous bianisotropic metasurface whose constitutive relations are given by~\cite{kong1986}
\begin{equation}
	\label{eq_Cdip}
	\begin{bmatrix}
		\ve{D}\\
		\ve{B}
	\end{bmatrix}=
	\begin{bmatrix}
		\te{\epsilon} & \te{\xi}\\
		\te{\zeta} & \te{\mu}
	\end{bmatrix}\cdot
	\begin{bmatrix}
		\ve{E}\\
		\ve{H}
	\end{bmatrix},
\end{equation}
where $\te{\epsilon}$ is the permittivity matrix, $\te{\mu}$ is the permeability matrix, and $\te{\xi}$ and $\te{\zeta}$ are magnetoelectro-coupling matrices. In what follows, we will consider that this metasurface is reciprocal, implying that 
\begin{equation}
	\label{eq_reci}
	\te{\epsilon} = \te{\epsilon}^T,\quad \te{\mu} = \te{\mu}^T,\quad \te{\zeta} = -\te{\xi}^T,	
\end{equation}
where $T$ is the transpose operation~\cite{kong1986}.

In order to obtain the invariance conditions that apply to parameters $\te{\epsilon}$, $\te{\mu}$, $\te{\xi}$ and $\te{\zeta}$ in~\eqref{eq_Cdip}, we shall first understand how the electric and magnetic fields, $\ve{E}$ and $\ve{H}$, transform under a given symmetry operation. For this purpose, consider the transformation matrix $\te{\Lambda}$ that corresponds to an arbitrary symmetry operation such as those described in App.~\ref{appendix}. It follows that $\ve{E}$ and $\ve{H}$ respectively transform into $\ve{E}'$ and $\ve{H}'$ as~\cite{arnaut1997,dmitriev2000}
\begin{subequations}
	\label{eq_TransEH}
	\begin{align}
		\ve{E}' &= \te{\Lambda}\cdot\ve{E},\label{eq_TransEH1}\\
		\ve{H}' &= \frac{1}{j\omega\epsilon_0}\left(\te{\Lambda}\cdot\nabla\right)\times\left(\te{\Lambda}\cdot\ve{E}\right) =  \text{det}\left(\te{\Lambda}\right)\te{\Lambda}\cdot\ve{H}\label{eq_TransEH2},
	\end{align}
\end{subequations}
where we have used the fact that the magnetic field $\ve{H}$, being a pseudovector, transforms as the curl of electric field $\ve{E}$. Note that all polar vectors would transform in the same way as $\ve{E}$ in~\eqref{eq_TransEH1}, whereas all pseudovector vectors transform as $\ve{H}$ in~\eqref{eq_TransEH2}. For the system in~\eqref{eq_Cdip}, we thus have that
\begin{subequations}
	\label{eq_TransDB}
	\begin{align}
		\ve{D}' &= \te{\Lambda}\cdot\ve{D},\\
		\ve{B}' &= \text{det}\left(\te{\Lambda}\right)\te{\Lambda}\cdot\ve{B},
	\end{align}
\end{subequations}

We next use~\eqref{eq_TransEH} and~\eqref{eq_TransDB} to obtain the transformation relations that apply to the material parameters in~\eqref{eq_Cdip}. To do so, we reverse~\eqref{eq_TransEH} and~\eqref{eq_TransDB} to express $\ve{E}$, $\ve{H}$, $\ve{D}$ and $\ve{B}$ in terms of $\ve{E}'$, $\ve{H}'$, $\ve{D}'$ and $\ve{B}'$, respectively, and then substitute the resulting relations into~\eqref{eq_Cdip}. By association, this readily yields~\cite{arnaut1997}
\begin{subequations}
	\label{eq_dtrans}
	\begin{align}
		\te{\epsilon}'&=\te{\Lambda}\cdot\te{\epsilon}\cdot\te{\Lambda}^{T},\\
		\te{\mu}'&=\te{\Lambda}\cdot\te{\mu}\cdot\te{\Lambda}^{T},\\
		\te{\xi}'&=\text{det}\left(\te{\Lambda}\right)\te{\Lambda}\cdot\te{\xi}\cdot\te{\Lambda}^{T},\\
		\te{\zeta}'&=\text{det}\left(\te{\Lambda}\right)\te{\Lambda}\cdot\te{\zeta}\cdot\te{\Lambda}^{T},
	\end{align}
\end{subequations}
where we have used the fact that $\te{\Lambda}$ is an orthogonal matrix implying that $\te{\Lambda}^{-1} = \te{\Lambda}^T$ and that $\text{det}(\te{\Lambda}) = \text{det}(\te{\Lambda})^{-1}$ since $\text{det}(\te{\Lambda}) = +1$ for rotation symmetries and $\text{det}(\te{\Lambda}) = -1$ for reflection symmetries (refer to App.~\ref{sec_Appsym}). 

The relations in~\eqref{eq_dtrans} represent how dipolar material parameters change under a particular spatial transformation defined by $\te{\Lambda}$. We shall see in Sec.~\ref{sec_inv} how such relations may be transformed into symmetry invariance conditions but, first, we shall investigate how these relations may be extended to an arbitrary multipolar order, which is the topic of the next section.

\subsection{Extension to multipolar responses}
\label{sec_mult}

As we shall see in Sec.~\ref{sec_examples}, some electromagnetic effects cannot be described solely using a purely dipolar model. This motivates the need to extend the dipolar framework discussed in the previous section to include higher-order multipolar components and their associated spatially-dispersive responses. For this purpose, we combine concepts from the multipolar theory typically used for modeling metamaterials~\cite{graham1990,raab2005,muhlig2011,alaee2019,evlyukhin2019} along with concepts associated to spatial dispersion~\cite{tretyakov2001,agranovich2014,simovski2018}. For the sake of conciseness, we shall next restrict ourselves to dipolar and quadrupolar responses, while higher-order multipolar responses may be considered in future works by following a procedure identical to the one discussed thereafter.

It follows that the quadrupolar constitutive relations read~\cite{simovski2018,achouri2021c,achouri2021b}
\begin{subequations}
	\label{eq_DBq}
	\begin{align}
		\ve{D} &= \epsilon_0\ve{E} + \ve{P} + \nabla\cdot\te{Q},\\
		\ve{B} &= \mu_0\left(\ve{H} + \ve{M} +  \nabla\cdot\te{S}\right),
	\end{align}
\end{subequations}
where $\ve{P}$ and $\ve{M}$ are electric and magnetic polarization densities, and  $\te{Q}$ and $\te{S}$ are irreducible (symmetric and traceless) electric and magnetic quadrupolar density tensors, respectively. The reason for considering irreducible tensors is that they provide a physically consistent description of the metasurface response~\cite{nanz2016}. These quantities may be related to the fields via the spatially dispersive relations~\cite{achouri2021c,achouri2021b}
\begin{equation}
	\label{eq_Qdip}
	\begin{bmatrix}
		P_i\\
		M_i\\
		Q_{il}\\
		S_{il}
	\end{bmatrix}\propto
	\begin{bmatrix}
		\chi_{\text{ee}}^{ij} & \chi_{\text{em}}^{ij} & \chi_{\text{ee}}^{'ijk} & \chi_{\text{em}}^{'ijk}\\
		\chi_{\text{me}}^{ij} & \chi_{\text{mm}}^{ij} & \chi_{\text{me}}^{'ijk} & \chi_{\text{mm}}^{'ijk}\\
		Q_{\text{ee}}^{ilj} & Q_{\text{em}}^{ilj} & Q_{\text{ee}}^{'iljk} & Q_{\text{em}}^{'iljk}\\
		S_{\text{me}}^{ilj} & S_{\text{mm}}^{ilj} & S_{\text{me}}^{'iljk} & S_{\text{mm}}^{'iljk}
	\end{bmatrix}\cdot
	\begin{bmatrix}
		E_{j}\\
		H_{j}\\
		\nabla_k E_{j}\\
		\nabla_k H_{j}
	\end{bmatrix},
\end{equation}
where $\chi_{\text{ee}}^{ij}$, $\chi_{\text{em}}^{ij}$, $\chi_{\text{me}}^{ij}$ and $\chi_{\text{mm}}^{ij}$ are related to the parameters in~\eqref{eq_Cdip}, whereas all the other terms are there to fully model the quadrupolar response of the metasurface. Note that reciprocity places relationships between parameters in~\eqref{eq_Qdip}, with the relevant reciprocity relations presented in~\cite{achouri2021c}.

As done in~\eqref{eq_dtrans}, we may express the transformation relations that apply to the parameters in~\eqref{eq_Qdip}. To do so, we make use of tensor notation\footnote{In what follows, we omit the summations over repeated indices for convenience.} since it applies more conveniently to the third and fourth order tensorial parameters in~\eqref{eq_Qdip}. It follows that an arbitrary tensor $\te{T}$, corresponding to any of the tensorial parameters of order 1 to 4 in~\eqref{eq_Qdip}, transforms under the transformation $\te{\Lambda}$ as~\cite{thompson1994}
\begin{subequations}
	\label{eq_TransT}
	\begin{align}
		T'_{i} &= a\Lambda_{ij}T_{j},\label{eq_TransT1}\\
		T'_{ij} &= a\Lambda_{im}\Lambda_{jk}T_{mk},\label{eq_TransT2}\\
		T'_{ijk} &= a\Lambda_{il}\Lambda_{jm}\Lambda_{kn}T_{lmn},\label{eq_TransT3}\\
		T'_{ijkl} &= a\Lambda_{im}\Lambda_{jn}\Lambda_{ko}\Lambda_{lp}T_{mnop},\label{eq_TransT4}
	\end{align}
\end{subequations}
where $a = \text{det}(\te{\Lambda})^n$ with $n=0$ for the `ee' or `mm' tensors in~\eqref{eq_Qdip} and $n=1$ for the `em' or `me' tensors\footnote{In the case of polar vectors, $n=0$, whereas $n=1$ for pseudovectors.}, respectively. It is clear that~\eqref{eq_TransT1} and~\eqref{eq_TransT2} are the tensor notation counterparts of~\eqref{eq_TransEH} and~\eqref{eq_dtrans}, respectively.

\subsection{Invariance conditions for material tensors}
\label{sec_inv}

We have just established that under a given symmetry operation, the material parameters in~\eqref{eq_Qdip} transform according to the relations~\eqref{eq_TransT}. We are now interested in connecting the material parameters in~\eqref{eq_Qdip} to the spatial symmetries of a metasurface. This may be achieved by considering that, according to the Neumann's principle, if a given structure is \emph{invariant} under a symmetry operation, then its material parameters should also be \emph{invariant} under the same operation~\cite{voigt1910,post1978,oliveira2008}. Such an invariance condition may be mathematically expressed from~\eqref{eq_TransT} as
\begin{subequations}
	\label{eq_InvCondT}
	\begin{align}
		T_{i} &= a\Lambda_{ij}T_{j},\\
		T_{ij} &= a\Lambda_{im}\Lambda_{jk}T_{mk},\label{eq_InvCondT2}\\
		T_{ijk} &= a\Lambda_{il}\Lambda_{jm}\Lambda_{kn}T_{lmn},\\
		T_{ijkl} &= a\Lambda_{im}\Lambda_{jn}\Lambda_{ko}\Lambda_{lp}T_{mnop},
	\end{align}
\end{subequations}
which implies that the tensor $\te{T}$ remains equal to itself after being transformed by $\te{\Lambda}$.

In order to obtain the material parameters that correspond to a given metasurface structure, we shall now describe a technique directly based on~\eqref{eq_InvCondT}. This technique differs from those described in the literature such as those that consist in expressing the material parameters for various symmetry groups~\cite{arnaut1997,dmitriev2000} or those using the orthogonality theorem to find the irreducible representation of the structure~\cite{baena2007,padilla2007,isik2009,yu2019}. Instead, we propose an approach that consists in recursively solving~\eqref{eq_InvCondT} for each symmetry operation for which the metasurface structure is invariant. By starting with a full material parameter tensor $\te{T}$ (one that contains all parameters), each iteration of this procedure leads to a system of equations formed by~\eqref{eq_InvCondT} that, when solved, reduces the complexity of $\te{T}$ by connecting some of its components together or by setting others to zero. At the end of this process, one is left with a tensor that is precisely invariant under all symmetry operations that define the metasurface structure and thus properly models its effective response.

\begin{figure}[h!]
	\centering
	\subfloat[]{\label{fig_unit1}
		\includegraphics[width=0.4\columnwidth]{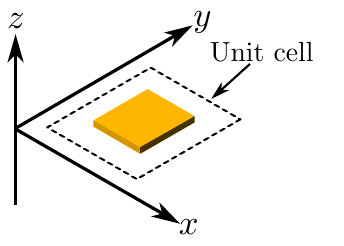}
	}\quad
	\subfloat[]{\label{fig_unit2}
		\includegraphics[width=0.4\columnwidth]{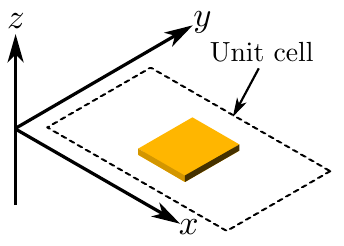}
	}
	\caption{Two different metasurface unit cells made of identical scattering particles arranged within (a) a square lattice and (b) a rectangular lattice.}
	\label{fig_unit}
\end{figure}

To illustrate this process, let us consider two different metasurfaces formed by periodically arranging in the $xy$-plane either the unit cell shown in Fig.~\ref{fig_unit1} or the one shown in Fig.~\ref{fig_unit2}.
These two unit cells are composed of the same scattering particle and only differ in the geometry of their lattice.

Inspecting the symmetries of these two metasurfaces, it is obvious that both are invariant under reflections through the $x$, $y$ and $z$ axes, which corresponds to the symmetry operations $\te{\Lambda}= \sigma_x, \sigma_y, \sigma_z$ (see  App.~\ref{sec_Appsym}). However, although they are composed of the same scattering particle, the metasurface unit cell in Fig.~\ref{fig_unit1} possesses a square lattice, whereas the one in Fig.~\ref{fig_unit2} possesses a rectangular lattice. It follows that these two metasurfaces do not exhibit the same rotation symmetry along the $z$-axis. Indeed, a metasurface composed of a periodic arrangement of the unit cell in Fig.~\ref{fig_unit1} would have a $\te{\Lambda}=C_{4,z}$ rotation symmetry (corresponding to $90^\circ$-rotation symmetry along $z$, see~\eqref{eq_Cni}), whereas the one composed of the unit cell in Fig.~\ref{fig_unit2} would have a $\te{\Lambda}=C_{2,z}$ rotation symmetry (corresponding to $180^\circ$-rotation symmetry along $z$).

Now that we have found the symmetries corresponding to the two metasurfaces in Fig.~\ref{fig_unit}, we can apply the process described above to find the permittivity matrix of these structures. Here, we restrict ourselves to finding $\te{\epsilon}$ for simplicity but without loss of generality since the process to obtain the other material parameters in~\eqref{eq_Qdip} would be identical. We provide more complete examples that include all material parameters in Sec.~\ref{sec_examples}.

The process to find the permittivity matrix corresponding to the metasurfaces in Fig.~\ref{fig_unit} starts by considering the full permittivity matrix given by
\begin{equation}
	\label{eq_fulleps}
	\te{\epsilon} =
	\begin{pmatrix}
		\epsilon_{xx} & \epsilon_{xy} & \epsilon_{xz}\\
		\epsilon_{xy}& \epsilon_{yy} & \epsilon_{yz}\\
		\epsilon_{xz} & \epsilon_{yz} & \epsilon_{zz}
	\end{pmatrix},
\end{equation}
where the reciprocity conditions~\eqref{eq_reci} have been considered. Next, relation~\eqref{eq_InvCondT2} is solved using~\eqref{eq_fulleps} and $\te{\Lambda}= \sigma_x$ with $\sigma_x$ given in~\eqref{eq_px}, which leads to the simplified permittivity matrix
\begin{equation}
	\label{eq_epspx}
	\te{\epsilon} =
	\begin{pmatrix}
		\epsilon_{xx} & 0 & 0\\
		0& \epsilon_{yy} &  \epsilon_{yz}\\
		0 &  \epsilon_{yz} & \epsilon_{zz}
	\end{pmatrix}.
\end{equation}
This process is then repeated by solving again~\eqref{eq_InvCondT2} but this time with~\eqref{eq_epspx} and $\te{\Lambda}= \sigma_y$. And then once again with $\te{\Lambda}= \sigma_z$ to obtain
\begin{equation}
	\label{eq_epspxpypz}
	\te{\epsilon} =
	\begin{pmatrix}
		\epsilon_{xx} & 0 & 0\\
		0& \epsilon_{yy} & 0\\
		0 & 0 & \epsilon_{zz}
	\end{pmatrix}.
\end{equation}
We are now left with the two rotation symmetries $C_{4,z}$ and $C_{2,z}$. However, it turns out that the permittivity matrix in~\eqref{eq_epspxpypz} already corresponds to the structure in Fig.~\ref{fig_unit2}. Indeed, we do not even need to apply a further iteration of the process with $\te{\Lambda}=C_{2,z}$ because a structure possessing reflection symmetries along the $x$ and $y$ axes \emph{necessarily} and \emph{automatically} possesses a $C_{2,z}$ rotation symmetry\footnote{However, the opposite is not necessarily true. Indeed, a structure possessing a $C_{2,z}$ symmetry, such as an S-shaped structure lying in the $xz$-plane, would not possess a reflection symmetry along the $x$-axis.}. This would therefore make a further application of the process with $\te{\Lambda}=C_{2,z}$ redundant. On the other hand, the rotation symmetry $C_{4,z}$ is not redundant and applying~\eqref{eq_InvCondT2} with $\te{\Lambda}=C_{4,z}$ and~\eqref{eq_epspxpypz} leads to 
\begin{equation}
	\label{eq_epspxpypzc4z}
	\te{\epsilon} =
	\begin{pmatrix}
		\epsilon_{xx} & 0 & 0\\
		0& \epsilon_{xx} & 0\\
		0 & 0 & \epsilon_{zz}
	\end{pmatrix},
\end{equation}
which is the permittivity matrix that corresponds to the metasurface in Fig.~\ref{fig_unit1}.

Obviously, for such simple structures as those in Fig.~\ref{fig_unit}, the results obtained in~\eqref{eq_epspxpypz} and~\eqref{eq_epspxpypzc4z} could have been guessed simply based on intuition. However, for more complex structures, intuition alone is usually not sufficient to guess the correct shape of the material parameter tensors, especially those related to quadrupolar responses in~\eqref{eq_Qdip}. The usefulness of the approach in such cases is emphasized using examples later in Sec.~\ref{sec_examples}.   

Note that the background medium is assumed to be identical on both sides of the metasurfaces in Fig.~\ref{fig_unit}. However, in practice, the scattering particles are usually deposited on top of a substrate instead of being embedded into a uniform background medium. In this case, the presence of the substrate would actually break the symmetry of the system in the $z$-direction meaning that the metasurface response would not be invariant under $\te{\Lambda} = \sigma_z$. While this would not change the shape of the permittivity matrices~\eqref{eq_epspxpypz} and~\eqref{eq_epspxpypzc4z}, it would have significant influence on the other material parameters in~\eqref{eq_Qdip}. For instance, it has been shown that the presence of a substrate is sufficient to lead to 3D chiral responses for metasurfaces composed of only 2D chiral scattering particles such as Gammadion structures~\cite{arteaga2016}.

\section{Scattering Matrix from Symmetries}
\label{sec_Appsym}

In the previous sections, we have seen how the effective material tensors of a metasurface may be predicted by considering the spatial symmetries of the metasurface unit cell. We shall now investigate the relationships between the scattering properties of a metasurface and its spatial symmetries. 

It turns out that the spatial symmetries of the metasurface unit cell are not the only parameters to consider when investigating the metasurface scattering response. Indeed, one needs to also take into account the influence of the waves interacting with the metasurface. This is because the combined system composed of the superposition of the metasurface and the incident and scattered waves exhibits spatial symmetries that are not the same of those of the metasurface alone~\cite{li2000,kahnert2005,dmitriev2011,dmitriev2013a,arteaga2014,kruk2015a,kruk2020}. Such considerations have already been well discussed in the literature, specifically in the pioneering works by Dmitriev for the case of normally incident waves~\cite{dmitriev2011} and oblique propagation~\cite{dmitriev2013a}.

In what follows, we shall review the general concepts discussed in~\cite{dmitriev2011,dmitriev2013a} for the case of reciprocal metasurfaces and develop a procedure to obtain the scattering matrix of a metasurface for a given direction of propagation that is similar to the procedure described in Sec.~\ref{sec_inv}.

\begin{figure}[h!]
	\centering
	\includegraphics[width=0.6\columnwidth]{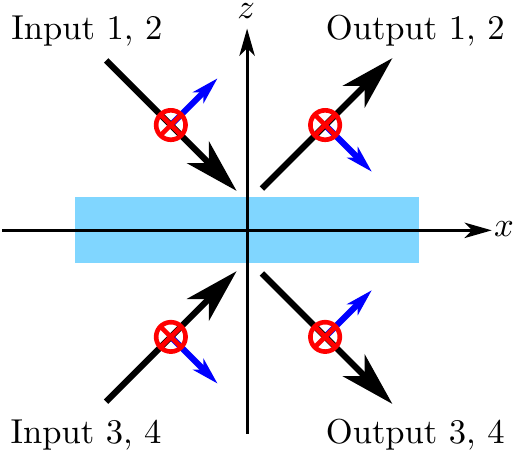}
	\caption{Cross-sectional view of a metasurface interacting with TE and TM polarized plane waves.}
	\label{fig_S}
\end{figure}

Let us first consider the case of a reciprocal metasurface illuminated by obliquely propagating plane waves. Since the metasurface is considered to be uniform and its lattice period is deeply subwavelength, it scatters the incident waves according to Snell's law. Such a situation is depicted in Fig.~\ref{fig_S}, where
two input and output ``ports'' transmit and receive, respectively, TE and TM polarized waves. Here, the plane of incidence is arbitrarily chosen to be the $xz$-plane. The scattering matrix associated with the interactions in Fig.~\ref{fig_S} is given by~\cite{dmitriev2011,dmitriev2013a} 
\begin{equation}
	\label{eq_S}
	\te{S} =
	\begin{bmatrix}
		R_{11} & R_{12} & T_{13} & T_{14} \\
		R_{21} & R_{22} & T_{23} & T_{24} \\
		T_{31} & T_{32} & R_{33} & R_{34} \\
		T_{41} & T_{42} & R_{43} & R_{44}
	\end{bmatrix},
\end{equation}
where $R_{ab}$ and $T_{ab}$ are reflection and transmission coefficients, respectively, with $a,b=1,2,3,4$ corresponding to the polarization state of the different waves where even and odd numbers are associated to TE and TM polarizations, respectively.

We are now looking for an invariance condition that would apply to $\te{S}$ in a similar way that the invariance conditions in~\eqref{eq_InvCondT} applied to the material tensor $\te{T}$. For this purpose, we need to relate $\te{S}$ to the symmetries of the combined system (metasurface and direction of wave propagation) defined by $\te{\Lambda}$. In that prospect, we start by defining a rotated version of $\te{\Lambda}$ that is aligned with the orientation of the plane of incidence since the latter is generally not restricted to the $xz$-plane. This rotated version of $\te{\Lambda}$ is given by
\begin{equation}
	\label{eq_RL}
	\te{\Lambda}' =  {R}_z(\phi)\cdot\te{\Lambda}\cdot{R}_z(\phi)^T,
\end{equation}
where the rotation matrix $R_z$ is defined in~\eqref{eq_rz} and $\phi$ represents the angle between the plane of incidence and the $xz$-plane. Consequently, if the plane of incidence coincides with the $xz$-plane, then $\phi=0^\circ$ or $\phi=180^\circ$, whereas if it is aligned with the $yz$-plane, then $\phi=90^\circ$ or $\phi=270^\circ$.

Now, since we are considering a given plane of incidence, it follows that only the spatial symmetries that map an input/output port to another one are to be considered for obtaining a symmetry invariant scattering matrix~\cite{dmitriev2013a}. The other symmetry operations, such as the $C_{4,z}$ rotation symmetry, should not be considered since they would be inconsistent with a fixed the plane of incidence. Therefore, the only spatial symmetries that make sense to be taken into account are $\te{\Lambda} =\sigma_i$ and $\te{\Lambda} =C_{2,i}$ where $i=x,y,z$ as well as the inversion symmetry operation defined by $\te{\Lambda} = -\te{I}$. 

Among this set of relevant symmetry operations, we may further classify them according to their effects on the polarization and the direction of wave propagation of the incident and scattered waves. For instance, applying the symmetry operation $\te{\Lambda} =\sigma_x$ on the system in Fig.~\ref{fig_S} would flip the sign of the TM polarizations, while leaving the TE polarizations unaffected, and it would also flip the direction of the propagation vectors $\ve{k}$. On the other hand, the operation $\te{\Lambda} =\sigma_y$ would flip the sign of the TE polarizations, while leaving both the TM polarizations and the $\ve{k}$-vectors unchanged.

Finally, we need to define a transformation operation that would apply to $\te{S}$ in the same way that the tensor $\te{T}$ was transformed by the operation $\te{\Lambda}$ in~\eqref{eq_TransT}. This may be achieved by defining the transformed scattering matrix, $\te{S}'$, as $\te{S}'=\te{M}\cdot\te{S}$, where $\te{M}$ is a transformation matrix that is related to $\te{\Lambda}$~\cite{dmitriev2013a}. Note that in~\cite{dmitriev2013a}, a specific transformation matrix $\te{M}$ was defined, for each of the relevant symmetry operations specified previously, based on the intuitive effects that these operations have on the polarizations and propagation directions of the waves. In what follows, we shall instead provide a technique that directly connects $\te{M}$ with $\te{\Lambda}$.

Noting that the invariance condition may be expressed as $\te{S}=\te{M}\cdot\te{S}\cdot\te{M}^{-1}$ for the symmetry operations that do not affect the direction of the $\ve{k}$-vectors, whereas it is given by $\te{S}=\te{M}\cdot\te{S}^T\cdot\te{M}^{-1}$ for the symmetry operations that flip the $\ve{k}$-vectors~\cite{dmitriev2013a}, we define the general invariance condition as
\begin{equation}
	\label{eq_InvS}
	\te{S}=\te{M}\cdot \left[ b\te{S} + (1-b)\te{S}^T  \right] \cdot\te{M}^{-1},
\end{equation}
where $b=1-\frac{c}{2}\left(1-\Lambda'_{xx}\right)$, with $\Lambda'_{xx} = \ve{x}\cdot\te{\Lambda}'\cdot\ve{x}$ and $c=c_1c_2$ and $c_1$ and $c_2$ given by
\begin{subequations}
	\begin{align}
		c_1 &= \begin{cases}
			1,~\text{if}~\Lambda'_{xz}=\Lambda'_{yz}=0 \\
			0,~\text{otherwise}
		\end{cases}\\	
		c_2 &= \begin{cases}
			1,~\text{if}~\te{\Lambda}'\cdot\ve{y}=\pm \te{R}_z(\phi)^T\cdot \te{R}_z(\phi)\cdot\ve{y} \\
			0,~\text{otherwise}
		\end{cases}
	\end{align}
\end{subequations}
The transformation matrix $\te{M}$ in~\eqref{eq_InvS} is now obtained by generalizing the results provided in~\cite{dmitriev2013a}, which leads to
\begin{equation}
	\label{eq_M}
	\te{M}=\begin{bmatrix}
		\te{a}_+ & \te{a}_-\\
		\te{a}_- & \te{a}_+
	\end{bmatrix},
\end{equation}
where the sub-matrices $\te{a}_+$ and $\te{a}_+$ are defined by
\begin{subequations}
	\label{eq_aa}
	\begin{align}
		\te{a}_+&=\frac{c}{2}\left(1+ \Lambda'_{zz}\right)\begin{bmatrix}
			\Lambda'_{yy} & 0 \\
			0 & \Lambda'_{xx}
		\end{bmatrix} + (1-c)\te{I}_\text{t},\\
		\te{a}_-&=\frac{c}{2}\left(1- \Lambda'_{zz}\right)\begin{bmatrix}
			\Lambda'_{yy} & 0 \\
			0 & \Lambda'_{xx}
		\end{bmatrix},
	\end{align}
\end{subequations}
where $\te{I}_\text{t}$ is a two-dimensional identity matrix.

The procedure to obtain the scattering matrix of a given metasurface under oblique incidence is quite similar to the approach described in Sec.~\ref{sec_inv} for finding the metasurface material parameter tensors. It consists in first defining the orientation of the plane of incidence with $\phi$, and then recursively solving~\eqref{eq_InvS} for each of the relevant symmetry operations expressed using~\eqref{eq_RL}.

For the special case of normally incident plane waves impinging on the metasurface, it is required to slightly modify this procedure~\cite{dmitriev2011}. First, at normal incidence, the orientation of the plane of incidence looses its meaning and~\eqref{eq_RL} should be bypassed by setting $\phi=0^\circ$. Then, the TE and TM polarizations should now be associated with $x$ and $y$ polarizations, respectively. Finally, it now actually makes sense to consider the $C_{4,z}$ rotation symmetry as it correctly maps TE to TM polarizations together. For this specific symmetry operation, the invariance condition~\eqref{eq_InvS} reduces to $\te{S}=\te{M}\cdot\te{S}\cdot\te{M}^{-1}$ with $\te{M}$ given by~\cite{dmitriev2011}
\begin{equation}
	\te{M}_{C_{4,z}} = 
	\begin{bmatrix}
		0 & -1 & 0 & 0	\\
		1 & 0 & 0 & 0	\\
		0 & 0 & 0 & -1	\\
		0 & 0 & 1 & 0	
	\end{bmatrix}.
\end{equation}
For all the other symmetry operations, relations~\eqref{eq_InvS} to~\eqref{eq_aa} should be used.

Before looking at examples illustrating the application of the method described above, which will be presented in the next section, we would like to comment on how to deduce the polarization effects induced by a metasurface directly from its scattering matrix. Considering the general scattering matrix~\eqref{eq_S}, we see that it may be reduced to 4 internal sub-matrices composed of two sets of reflection and transmission matrices. Each one of these sub-matrices may be associated to a Jones matrix since they describe how different polarizations are scattered by the metasurface~\cite{saleh2019}. From a given Jones matrix it is then possible to deduce the general effect that the metasurface has on different polarization states~\cite{menzel2010,kruk2015a,kruk2020,achouri2021a}. For completeness, we provide in Table.~\ref{tab_jones} the relationships between some common Jones matrices and their corresponding polarization effects.

\begin{table*}[h!]
	\footnotesize 	
	\begin{TAB}(r,1cm,1.5cm)[5pt]{c|c|c|c|c|c|c}{|c|c|}
		Jones matrix & 
		$\begin{bmatrix} 
			a & 0\\
			0 & a
		\end{bmatrix}$ & $\begin{bmatrix} 
			a & 0\\
			0 & d
		\end{bmatrix}$ & $\begin{bmatrix} 
			a & b\\
			b & d
		\end{bmatrix}$ & $\begin{bmatrix} 
			a & b\\
			-b & a
		\end{bmatrix}$ & $\begin{bmatrix} 
			a & b\\
			-b & d
		\end{bmatrix}$ & $\begin{bmatrix} 
			a & b\\
			c & d
		\end{bmatrix}$ \\ 
		Polarization effects    & None & LP biref.  & $\begin{matrix} 
			\text{LP biref.} \\
			\text{Pol. conversion} \\
			\text{Asym.transmission} 
		\end{matrix} $  & $\begin{matrix} 
			\text{CP biref.} \\
			\text{Pol. rotation} 
		\end{matrix} $ & $\begin{matrix} 
			\text{LP/CP biref.} \\
			\text{Pol. conversion} 
		\end{matrix}$ & $\begin{matrix} 
			\text{LP/CP biref.} \\
			\text{Pol. conversion} \\
			\text{Asym. transmission} 
		\end{matrix}$
	\end{TAB}
	\caption{\label{tab_jones}Jones matrices and their effects on the polarization state~\cite{menzel2010,kruk2015a,kruk2020,achouri2021a}. The terms ``LP biref.'' and ``CP biref.'' refer to linear and circular birefrigence, respectively, which also includes the effects related to linear and circular dichroism.}
\end{table*}

\section{Illustrative examples}
\label{sec_examples}

We shall now look at three examples that illustrate the application of the techniques described in Sec.~\ref{sec_mat} and in Sec.~\ref{sec_Appsym}. In these examples, we will represent the material parameter tensors in way that is reminiscent of how they are depicted in~\cite{bernalarango2014}. Meaning that we will make use of the fact that the quadrupolar tensors in~\eqref{eq_DBq} are assumed to be irreducible tensors (symmetric and traceless), which allows us to greatly reduce the number of independent components in~\eqref{eq_Qdip}. This also means that the derivative operators in~\eqref{eq_Qdip} can be simplify such that, for instance, the electric field gradient becomes~\cite{bernalarango2014}
\begin{equation}
	\label{eq_DE}
	\begin{split}
		\diamond\ve{E} &= \{(\partial_x E_y+\partial_y E_x)/2,(\partial_x E_z+\partial_z E_x)/2 \\
		&(\partial_y E_z+\partial_z E_y)/2, \partial_xE_x, \partial_yE_y, \partial_zE_z  \},
	\end{split}
\end{equation}
and similarly for the magnetic field gradient.

\subsection{Extrinsic chirality}
\label{sec_ec}

Extrinsic chirality corresponds to an electromagnetic effect where a medium (in our case a metasurface) exhibits a chiral response even though it is not composed of geometrically 3D chiral scattering particles~\cite{plum2009,plum2009a,cao2015,okamoto2019}. This chiral response emerges only when the metasurface is illuminated along a specific direction, hence the reason why it is referred to as ``extrinsic'': it depends on the direction of wave propagation.

\begin{figure}[h!]
	\centering
	\subfloat[]{\label{fig_T1}
		\includegraphics[width=0.4\columnwidth]{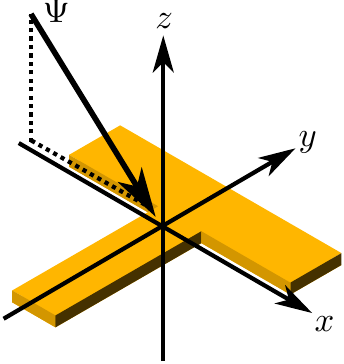}
	}\quad
	\subfloat[]{\label{fig_T2}
		\includegraphics[width=0.4\columnwidth]{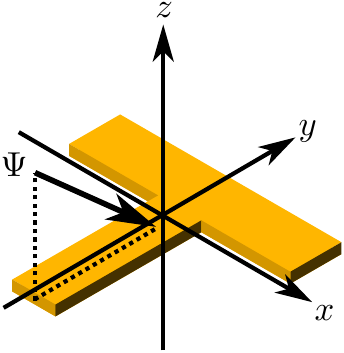}
	}
	\caption{Metasurface unit cell composed of a T-shaped particle being illuminated by an obliquely incident plane wave. The metasurface lattice is a square.}
	\label{fig_T}
\end{figure}

Consider the metasurface unit cell depicted in Fig.~\ref{fig_T} composed of a T-shaped metallic particle embedded in a uniform background medium. The metasurface is illuminated by an obliquely incident plane wave propagating either in the $xz$-plane, as in Fig.~\ref{fig_T1}, or in the $yz$-plane, as in Fig.~\ref{fig_T2}.

\begin{figure}[h!]	
	\centering
	\includegraphics[width=0.8\columnwidth]{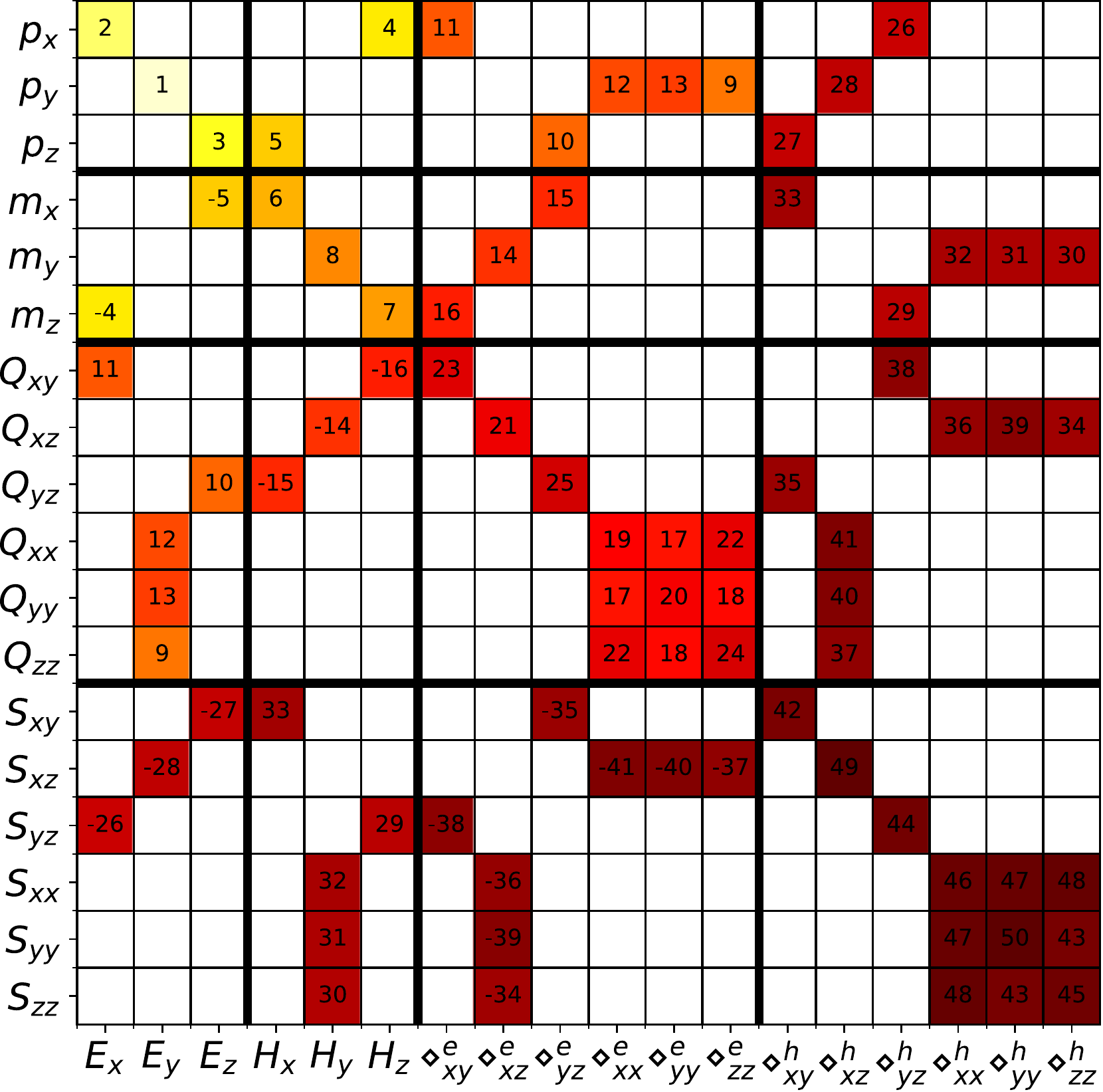}
	\caption{\label{fig_TX} Representation of the symmetry-allowed multipolar components corresponding to the metasurface in Fig.~\ref{fig_T}.}
\end{figure}

The spatial symmetries of the metasurface, assuming that it has a square lattice, are directly found to be $\sigma_x$ and $\sigma_z$. Note that it also exhibits a $C_{2,y}$ rotation symmetry, however, it is redundant since we have already considered its $\sigma_x$ reflection symmetry, as explained in Sec.~\ref{sec_inv}. Applying the procedure outlined in Sec.~\ref{sec_inv}, we find the multipolar components of this metasurface and present them in Fig.~\ref{fig_TX}, using~\eqref{eq_DE}, in way that is similar to the representations in~\cite{bernalarango2014}.

In Fig.~\ref{fig_TX}, the vertical and the horizontal axes respectively correspond to the multipolar densities and the excitation vector, on the left- and right-hand side of~\eqref{eq_Qdip}, where the term $\diamond_ {xy}^\text{e}$ corresponds to the first element in~\eqref{eq_DE} and so on. The numbers and the colors in the figure are purely arbitrary and only help visualize which components are allowed to exist due to the symmetries of the structure and which components are related to each other (components that have the same number and color).

The information provided in Fig.~\ref{fig_TX} reveals the complexity of the metasurface multipolar response, notably its bianisotropic dipolar nature due to the existence of the components numbered 4 and 5 in the matrix. Comparing the results in Fig.~\ref{fig_TX} with those of the dolmen structure provided in~\cite{bernalarango2014} reveals a perfect match between the two methods. It should be noted that results in Fig.~\ref{fig_TX} only correspond to the components that are allowed to exist and it does not provide any information about the magnitude of these components. On the other hand, the results presented in~\cite{bernalarango2014} provide the numerical value of the multipolar components since the method used to obtain them consists in numerically simulating the structure with a complex set of different illumination conditions, which results in an important computational cost.

Now, we use the approach described in Sec.~\ref{sec_mat} to compute the scattering matrix for the obliquely propagating waves of Fig.~\ref{fig_T}. Using~\eqref{eq_RL} to~\eqref{eq_aa} with $\te{\Lambda}=\sigma_x$, $\te{\Lambda}=\sigma_z$ and $\phi=0^\circ$ yields the scattering matrix, for the case in Fig.~\ref{fig_T1}, given by
\begin{equation}
	\label{eq_S_T_xz}
	\te{S} =
	\begin{bmatrix}
		R_{11} & R_{12} & T_{13} & T_{14} \\
		-R_{12} & R_{22} & -T_{14} & T_{24} \\
		T_{13} & T_{14} & R_{11} & R_{12} \\
		-T_{14} & T_{24} & -R_{12} & R_{22}
	\end{bmatrix}.
\end{equation}
Similarly, using $\phi=90^\circ$, we obtain the scattering matrix for the case in Fig.~\ref{fig_T2} as
\begin{equation}
	\label{eq_S_T_yz}
	\te{S} =
	\begin{bmatrix}
		R_{11} & 0 & T_{13} & 0 \\
		0 & R_{22} & 0 & T_{24} \\
		T_{13} & 0 & R_{11} & 0 \\
		0 & T_{24} & 0 & R_{22}
	\end{bmatrix}.
\end{equation}

Comparing~\eqref{eq_S_T_xz} and~\eqref{eq_S_T_yz}, it is clear that illuminating the metasurface along different direction of wave propagation leads to quite different polarization effects. When the metasurface is illuminated in the $xz$-plane, as in Fig.~\ref{fig_T1}, we see that the internal Jones matrices in~\eqref{eq_S_T_xz} exhibit circular birefringent properties typical of chiral responses (see Table~\ref{tab_jones}). On the other hand, illuminating the metasurface in the $yz$-plane, as in Fig.~\ref{fig_T2}, leads to a linear birefringent response, as evidenced by the scattering matrix in~\eqref{eq_S_T_yz}. 
This analysis confirms that the metasurface is extrinsically chiral since its chiral response is dependent on the direction of wave propagation.  

One way to understand the emergence of a chiral response in this structure is to consider its bianisotropic dipolar response. For an illumination in the $xz$-plane, both TE and TM polarizations excite the components 4 and 5 in Fig.~\ref{fig_TX}, which correspond to the components $\chi_{\text{em}}^{xz}$ and $\chi_{\text{em}}^{zx}$ in~\eqref{eq_Qdip}, respectively, or to their reciprocal counterparts $\chi_{\text{me}}^{zx}$ and $\chi_{\text{me}}^{xz}$. Since these two components have different values that thus cannot cancel each other, they lead to a net chiral response~\cite{chen2021}. However, for an illumination in the $yz$-plane, both TE and TM polarizations excite either the set of parameters $\chi_{\text{em}}^{xz}$ and $\chi_{\text{me}}^{zx}$ or the set $\chi_{\text{em}}^{zx}$ and $\chi_{\text{me}}^{xz}$, respectively. Since by reciprocity (see~\eqref{eq_reci}), $\chi_{\text{em}}^{xz}=-\chi_{\text{me}}^{zx}$ and $\chi_{\text{em}}^{zx}=-\chi_{\text{me}}^{xz}$, these two contributions cancel each other leading to an absence of chiral response.

\subsection{Asymmetric Angular Transmittance}

We are now interested in designing a metasurface that exhibits an asymmetric angular transmittance when illuminated in the $xz$-plane. While it has been shown that asymmetric angular transmittance may be achieved using spatially varying metasurfaces with phase-gradient modulations~\cite{wang2018}, we are instead interested in designing a \emph{uniform} metasurface whose asymmetric scattering response stems from the asymmetry of its structure.

\begin{figure}[h!]
	\centering
	\subfloat[]{\label{fig_L_flat}
		\includegraphics[width=0.4\columnwidth]{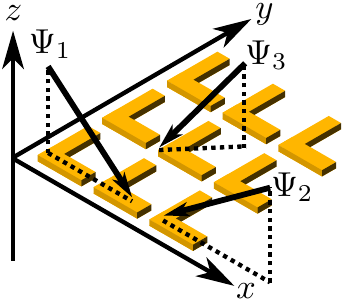}
	}\quad
	\subfloat[]{\label{fig_L_vert}
		\includegraphics[width=0.4\columnwidth]{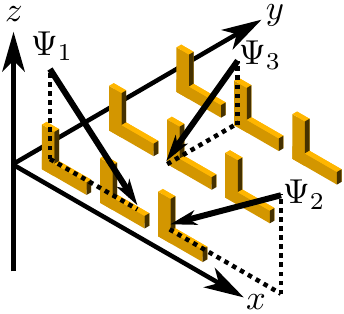}
	}
	\caption{Two metasurfaces composed of (a) planar L-shaped structures and (b) vertical L-shaped structures. In both cases, the structures have arms of equal length.}
	\label{fig_L}
\end{figure}

Let us consider the two metasurfaces depicted in Fig.~\ref{fig_L} composed of a periodic array of L-shaped scattering particles. In Fig.~\ref{fig_L_flat}, the structures lay flat on the plane of the surface, whereas they stand vertical in Fig.~\ref{fig_L_vert}.

Since both structures are asymmetric in the $x$-direction, one may a priori expect that they would both lead to an asymmetric scattering response for waves propagating in the $xz$-plane in opposite directions like the waves $\Psi_1$ and $\Psi_2$ in Fig.~\ref{fig_L}. However, that is not the case as we shall next demonstrate.

\begin{figure}[h!]
	\centering
	\subfloat[]{\includegraphics[width=0.8\columnwidth]{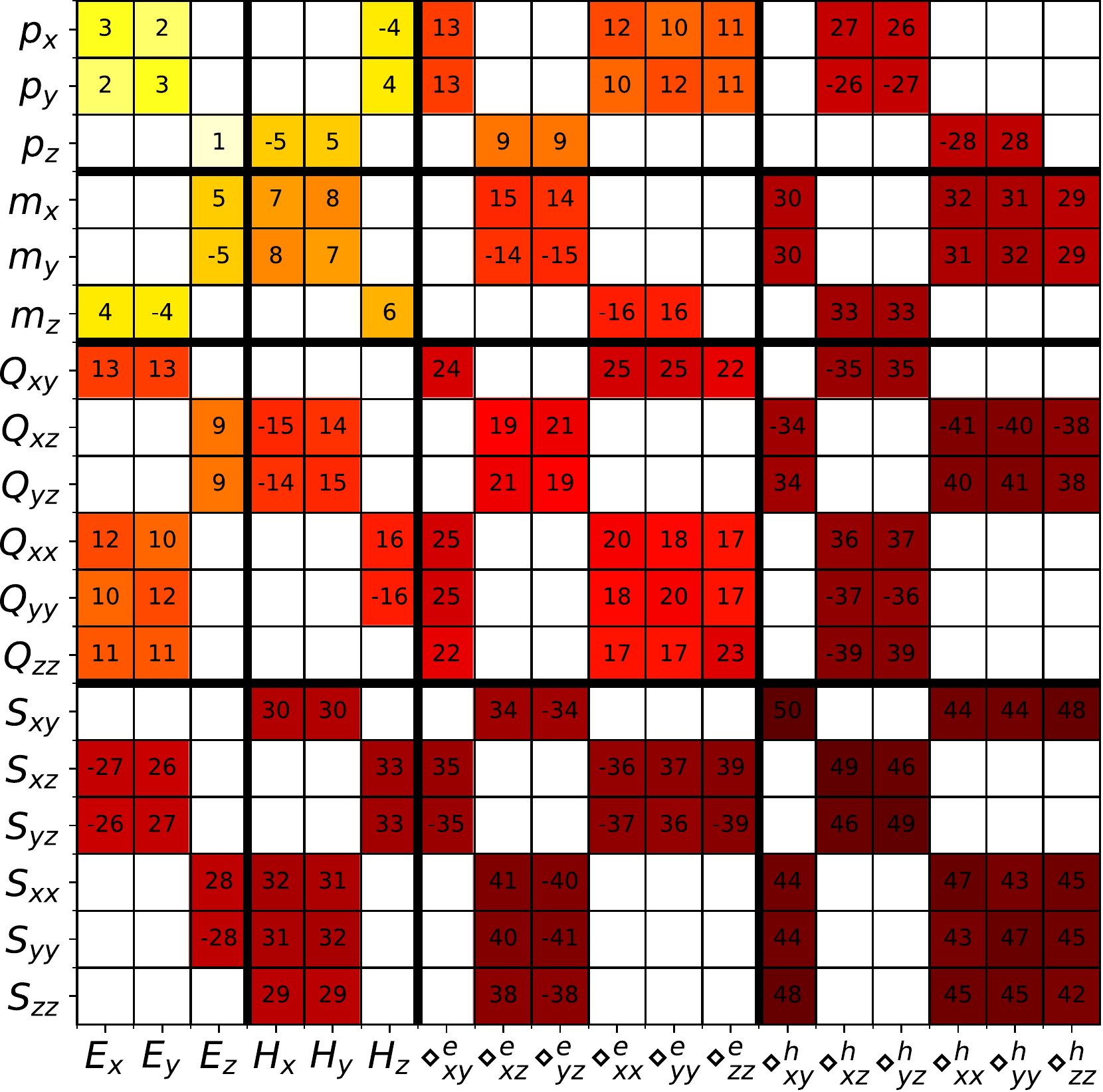}}\\
	\subfloat[]{\includegraphics[width=0.8\columnwidth]{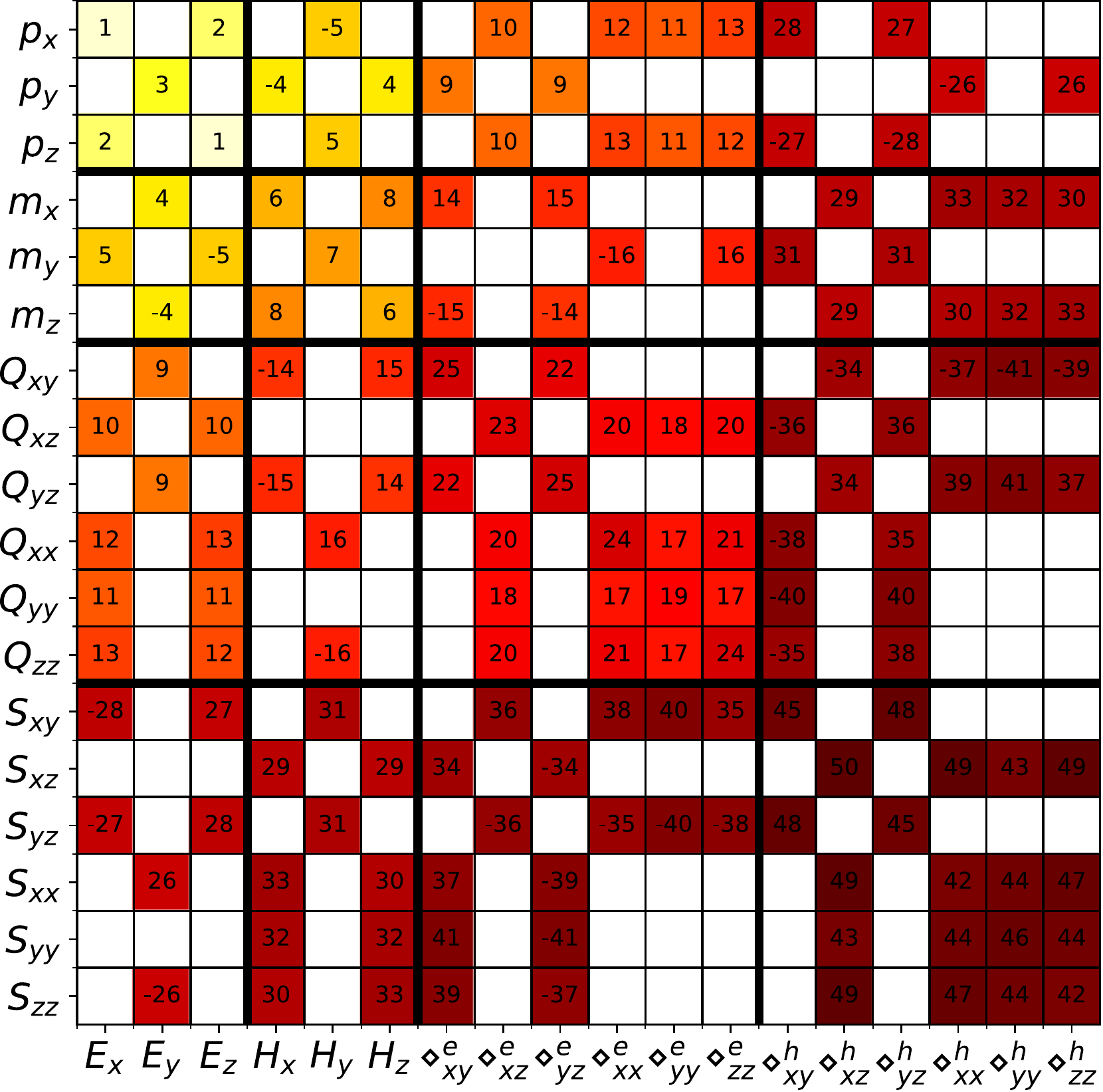}}
	\caption{\label{fig_MS_X} Symmetry-allowed material parameter tensors corresponding to the metasurface in (a)~Fig.~\ref{fig_L_flat} and in (b)~Fig.~\ref{fig_L_vert}.}
\end{figure}

The spatial symmetries of the structure in Fig.~\ref{fig_L_flat} are $\sigma_z$ and $\sigma_{xy}$, whereas those of the structure in Fig.~\ref{fig_L_vert} are $\sigma_y$ and $\sigma_{xz}$. Where $\sigma_{xy}$ and $\sigma_{xz}$ refer to symmetries with respect to the $45^\circ$-diagonal between the $x$- and $y$-axes or between the $x$- and $z$-axes, respectively. These symmetries may be defined by combining those provided in App.~\ref{appendix} so that $\sigma_{xy}=R_z(-45^\circ)\cdot\sigma_x\cdot R_z(45^\circ)$ and $\sigma_{xz}=R_y(-45^\circ)\cdot\sigma_x\cdot R_y(45^\circ)$. Applying the approach discussed in Sec.~\ref{sec_inv} to these two sets of symmetries yields their corresponding material parameter tensors that are provided in Fig.~\ref{fig_MS_X}. As can be seen, both structures are bianisotropic and, if one only considers their permittivity matrix, we see that $\epsilon_{xx}=\epsilon_{yy}$ and $\epsilon_{xy}\neq 0$ for the flat L-shaped structures of Fig.~\ref{fig_L_flat}, whereas $\epsilon_{xx}=\epsilon_{zz}$ and $\epsilon_{xz}\neq 0$ for the vertical L-shaped structures of Fig.~\ref{fig_L_vert}, as may be intuitively expected.

We next compute their respective scattering matrices. For the case in Fig.~\ref{fig_L_flat}, and the waves $\Psi_1$ and $\Psi_2$ that propagate in the $xz$-plane ($\phi=0^\circ$), we have that
\begin{equation}
	\label{eq_S_L_flat_xz}
	\te{S} =
	\begin{bmatrix}
		R_{11} & R_{12} & T_{13} & T_{14} \\
		-R_{12} & R_{22} & -T_{14} & T_{24} \\
		T_{13} & T_{14} & R_{11} & R_{12} \\
		-T_{14} & T_{24} & -R_{12} & R_{22}
	\end{bmatrix},
\end{equation}
Interestingly, we see that this structure exhibits a form of extrinsic chirality similarly to the one in Fig.~\ref{fig_T1}. However, this metasurface does not possess an asymmetric angular transmittance since the two transmission sub-matrices in~\eqref{eq_S_L_flat_xz} are equal to each other implying that $\Psi_1$ and $\Psi_2$ transmit through the metasurface identically.

Note that if the plane of incidence coincides with the in-plane axis of symmetry along which $\sigma_{xy}$ is satisfied, as it is the case for the wave $\Psi_3$ in Fig.~\ref{fig_L_flat}, the extrinsic chiral response disappears as evidenced by the scattering matrix ($\phi=45^\circ$)
\begin{equation}
	\label{eq_S_L_flat_xyz}
	\te{S} =
	\begin{bmatrix}
		R_{11} & 0 & T_{13} & 0 \\
		0 & R_{22} & 0 & T_{24} \\
		T_{13} & 0 & R_{11} & 0 \\
		0 & T_{24} & 0 & R_{22}
	\end{bmatrix}.
\end{equation}

For the metasurface in Fig.~\ref{fig_L_vert}, and the waves $\Psi_1$ and $\Psi_2$ that propagate in the $xz$-plane ($\phi=0^\circ$), we have that
\begin{equation}
	\label{eq_S_L_vert}
	\te{S} =
	\begin{bmatrix}
		R_{11} & 0 & T_{13} & 0 \\
		0 & R_{22} & 0 & T_{24} \\
		T_{31} & 0 & R_{33} & 0 \\
		0 & T_{42} & 0 & R_{44}
	\end{bmatrix}.
\end{equation}
This time, not only there is no cross-polarized scattering but the two transmission sub-matrices are different from each other ($T_{13}\neq T_{31}$ and $T_{24}\neq T_{42}$) implying that the waves $\Psi_1$ and $\Psi_2$ interact differently with the metasurface. This indicates that asymmetric angular transmittance is possible in this situation. For the incident wave $\Psi_3$ that propagates in the $yz$-plane, there is however no co-polarized angular transmittance asymmetry as evidence by its scattering matrix ($\phi=90^\circ$)
\begin{equation}
	\label{eq_S_L_vert2}
	\te{S} =
	\begin{bmatrix}
		R_{11} & R_{12} & T_{13} & T_{14} \\
		-R_{12} & R_{22} & T_{23} & T_{24} \\
		T_{13} & -T_{23} & R_{33} & R_{34} \\
		-T_{14} & T_{24} & -R_{34} & R_{44}
	\end{bmatrix}.
\end{equation}
Nevertheless illuminating the metasurface under this direction of propagation should induce a form of asymmetric transmittance when illuminated in the $+z$- or the $-z$-directions. This therefore does not correspond to the sought after asymmetric angular transmittance effect.

\subsection{Multipolar extrinsic chirality}

\begin{figure}[h!]
	\centering
	\subfloat[]{\label{fig_MS_square}
		\includegraphics[width=0.4\columnwidth]{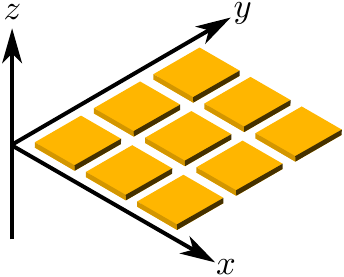}
	}\quad
	\subfloat[]{\label{fig_MS_gamm}
		\includegraphics[width=0.4\columnwidth]{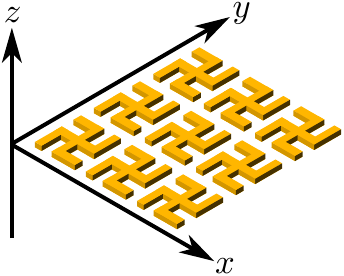}
	}
	\caption{Two metasurfaces composed of a periodic array of (a)~square particles and (b)~Gammadion particles.}
	\label{fig_MS}
\end{figure}

The last example investigates the scattering response of the two metasurfaces depicted in Fig.~\ref{fig_MS} under \emph{normal} illumination.
The metasurface in Fig.~\ref{fig_MS_square} consists of a simple periodic array of square-shaped scattering particles, whereas the one in Fig.~\ref{fig_MS_gamm} is made of Gammadion structures. In both cases, the metasurface is embedded within a uniform medium.  
\begin{figure}[h!]
	\centering
	\subfloat[]{\label{fig_MS_chiralX1}\includegraphics[width=0.8\columnwidth]{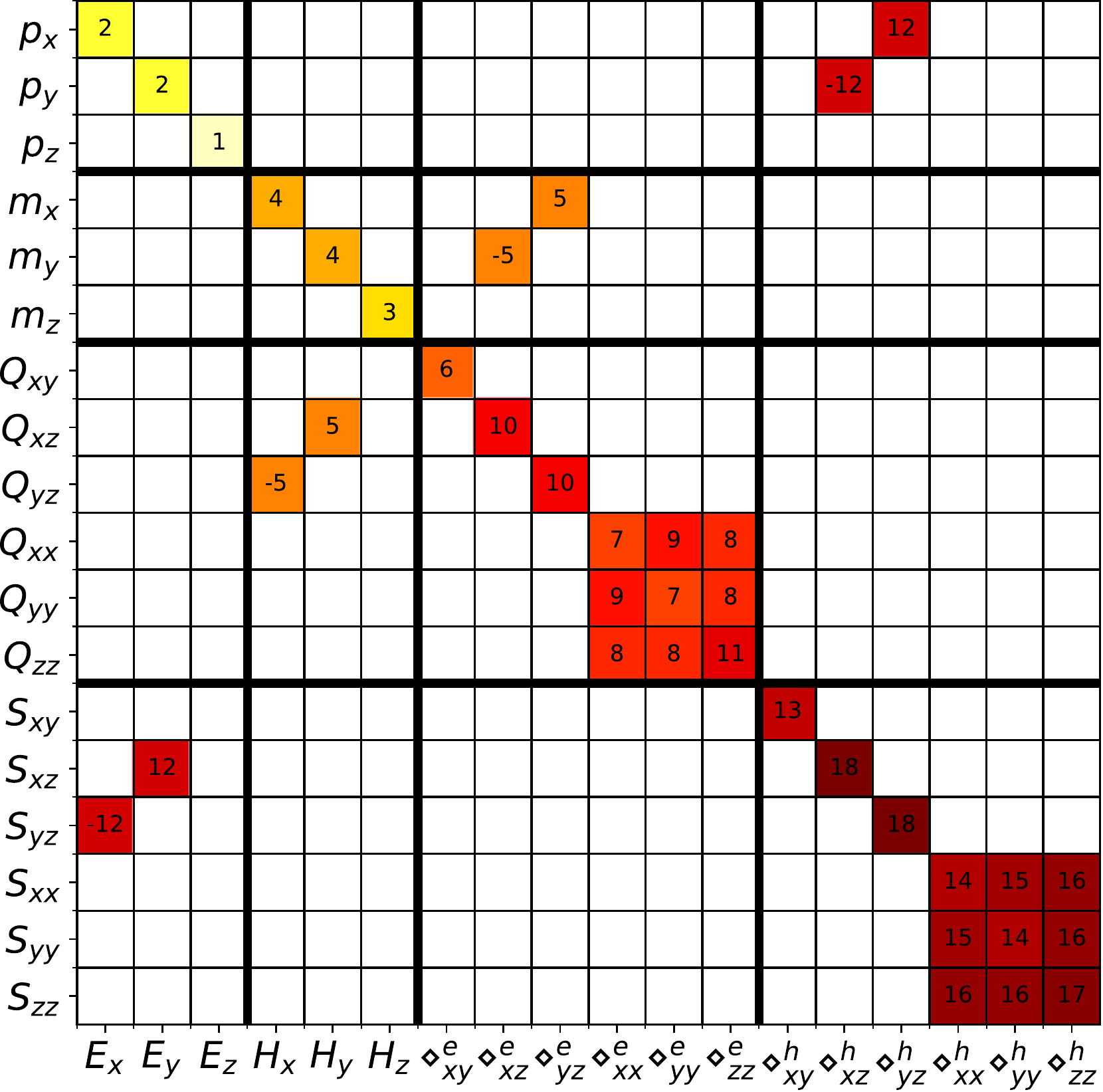}}\\
	\subfloat[]{\label{fig_MS_chiralX2}\includegraphics[width=0.8\columnwidth]{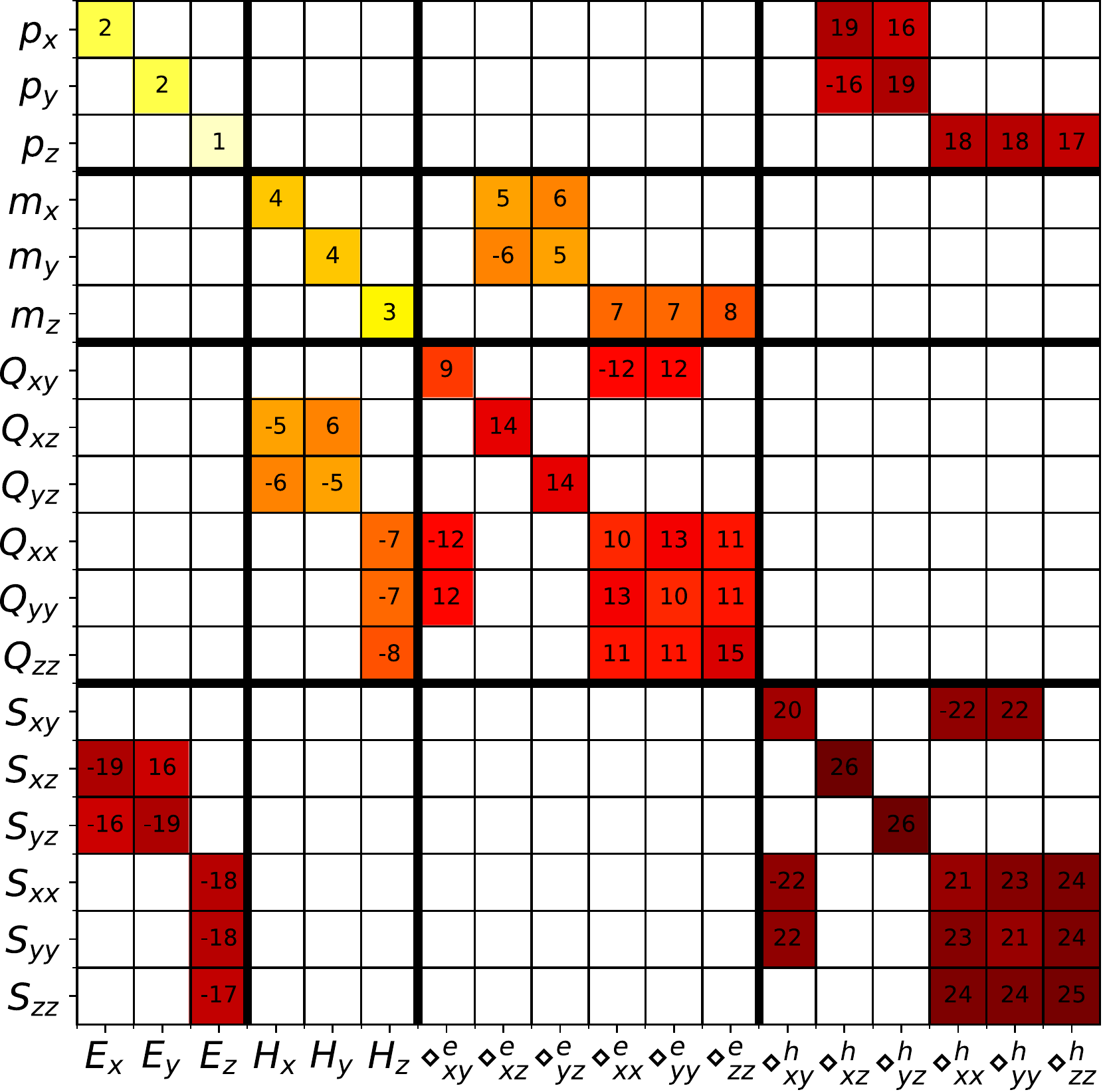}}
	\caption{\label{fig_MS_chiralX}Symmetry-allowed material parameter tensors corresponding to the metasurface in (a)~Fig.~\ref{fig_MS_square} and in (b)~Fig.~\ref{fig_MS_gamm}.}
\end{figure}
\begin{figure*}[t!]
	\centering
	\includegraphics[width=2\columnwidth]{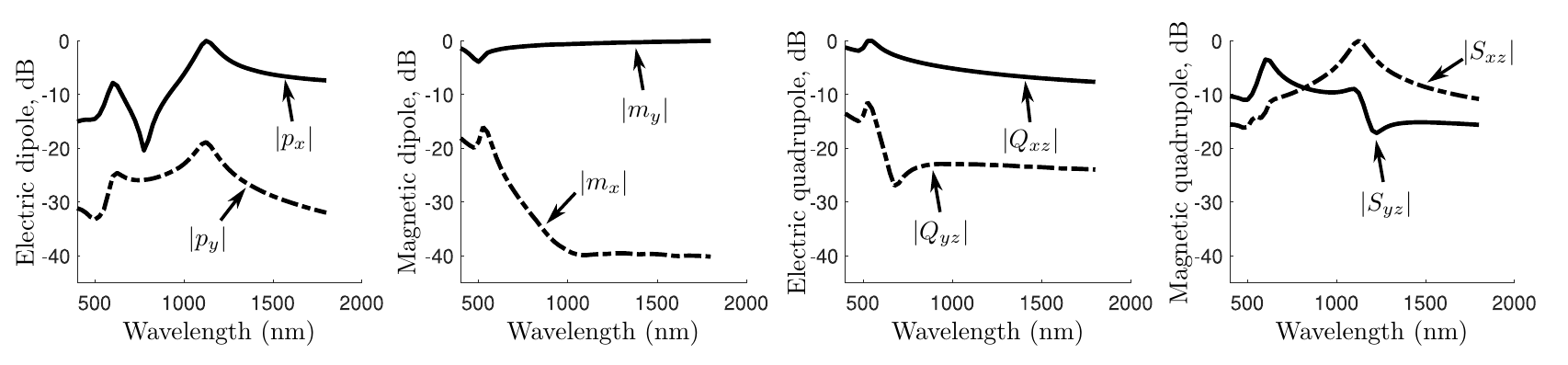}
	\caption{Simulated spherical dipolar and quadrupolar components of an isolated gold Gammadion structure illuminated by an $x$-polarized $z$-propagating plane wave. The arms of the Gammadion structure have a cross-section of $30\times30$~nm, while the footprint of the structure is $150\times150$~nm. The background medium is vacuum.}
	\label{fig_gam_sim}
\end{figure*}

While it is intuitive to guess that a normally incident plane wave impinging on the metasurface in Fig.~\ref{fig_MS_square} would not undergo polarization conversion, it is less trivial to intuitively predict the response of the metasurface in Fig.~\ref{fig_MS_gamm}. Referring to the literature, it was for instance suggested in~\cite{arteaga2016} that such an ideal Gammadion array should not exhibit a chiral response unless the scattering particles are asymmetric in the $z$-direction (or  the metasurface lays on a substrate). In what follows, we shall see that the problem is in fact more complex than it appears.

Let us first compute the material parameter tensors of these two metasurfaces. The symmetries associated with the metasurface in Fig.~\ref{fig_MS_square} are $\sigma_x$, $\sigma_y$, $\sigma_z$ and $C_{4,z}$, whereas those for the metasurface in Fig.~\ref{fig_MS_gamm} are only $\sigma_z$ and $C_{4,z}$. The corresponding material parameter tensors are given in Fig.~\ref{fig_MS_chiralX}.
Inspecting the differences between Fig.~\ref{fig_MS_chiralX1} and Fig.~\ref{fig_MS_chiralX2} reveals a striking feature: they both exhibit identical purely dipolar responses, i.e., the 4 $9\times9$ sub-matrices relating $\ve{p}$ and $\ve{m}$ to $\ve{E}$ and $\ve{H}$ are the same for both metasurfaces. This means that a modeling approach solely based on~\eqref{eq_Cdip} would fail to predict a difference in the scattering response of these metasurfaces. It is only by considering quadrupolar responses and higher-order spatially dispersive effects that the differences between these metasurface become evident.

Following the approach described in Sec.~\ref{sec_Appsym} for normally incident waves, the scattering matrix of the metasurface in Fig.~\ref{fig_MS_square} is given by
\begin{equation}
	\label{eq_S_MS_square}
	\te{S} =
	\begin{bmatrix}
		R_{11} & 0 & T_{13} &0 \\
		0 & R_{11} & 0 & T_{13} \\
		T_{13} & 0 & R_{11} & 0 \\
		0 & T_{13} & 0 & R_{11}
	\end{bmatrix},
\end{equation}
whereas the one for the metasurface in Fig.~\ref{fig_MS_gamm} reads
\begin{equation}
	\label{eq_S_MS_gamm}
	\te{S} =
	\begin{bmatrix}
		R_{11} & R_{12} & T_{13} & T_{14} \\
		-R_{12} & R_{11} & -T_{14} & T_{13} \\
		T_{13} & T_{14} & R_{11} & R_{12} \\
		-T_{14} & T_{13} & -R_{12} & R_{11}
	\end{bmatrix}.
\end{equation}
As may be expected, the metasurface in Fig.~\ref{fig_MS_square} is isotropic and does not induce any polarization rotation or conversion at normal incidence. On the other hand, the shape of the scattering matrix~\eqref{eq_S_MS_gamm} reveals that the metasurface in Fig.~\ref{fig_MS_gamm} should exhibit a chiral response. This may come as a surprise considering that the metasurface is not made of geometrically 3D chiral structures and that it is illuminated at normal incidence. However, a detailed inspection of the material parameter tensors in Fig.~\ref{fig_MS_chiralX2} helps understanding why such a chiral response exists.

Consider an $x$-polarized normally incident plane wave impinging on the metasurface in Fig.~\ref{fig_MS_gamm} with an electric field defined by $\ve{E}=\ve{\hat{x}}e^{i(\omega t-kz)}$ and a magnetic field given by $\ve{H}=\ve{\hat{y}}e^{i(\omega t-kz)}/\eta$ with $k$ and $\eta$ being the wavenumber and impedance of the background medium. This wave excites the components 5 and 19 in Fig.~\ref{fig_MS_chiralX2} via the field derivatives $\diamond_{xz}^\text{e}$ and $\diamond_{yz}^\text{h}$, which, in this case, correspond to $\partial_z E_x$ and $\partial_z H_y$, respectively. Since these spatial derivatives are not zero for the considered excitation, it follows that these two material components, which do not exist for the metasurface in Fig.~\ref{fig_MS_square}, respectively induce an $x$-polarized magnetization of the metasurface, $m_x$, and a $y$-polarized polarization, $p_y$, suggesting rotation of polarization. By reciprocity~\cite{achouri2021c}, the components 5 and 19 also induce the quadrupolar responses $Q_{yz}$ and $S_{xz}$ that are excited via the field components $H_y$ and $E_x$, respectively, and that also contribute to rotation of polarization.

In order to verify that the multipolar components $p_y$, $m_x$, $Q_{yz}$ and $S_{xz}$ are indeed excited when illuminating such a structure with an $x$-polarized normally propagating plane wave, we have numerically simulated an isolated Gammadion particle. From its scattered fields, we have extracted its spherical multipolar components following the approach provided in~\cite{alaee2018}. The resulting simulations are plotted in Fig.~\ref{fig_gam_sim}, where the solid and dashed lines correspond to the co- and cross-polarized multipolar components, respectively. It is thus clear that a cross-polarized response is achieved although being small compared to the co-polarized one. This confirms that the metasurface in Fig.~\ref{fig_MS_gamm} exhibits a chiral response as suggested by its scattering matrix~\eqref{eq_S_MS_gamm}. 

This type of chiral response slightly differs from the one discussed in Sec.~\ref{sec_ec}, where the chirality was due to an obliquely propagating wave interacting with the bianisotropic effective material parameters of the structure. In the case of the metasurface in Fig.~\ref{fig_MS_gamm}, the chiral response is due to multipolar and spatially-dispersive components excited by a normally propagating plane, thus indicating the presence of multipolar extrinsic chirality. The extrinsic nature of this chiral response is here not directly related to the direction of wave propagation, as it was the case in Sec.~\ref{sec_ec}, but rather to the gradient of the fields.

\section{Conclusions}

We have established a relationship between the spatial symmetries of a metasurface and its corresponding material parameter tensors and scattering matrix. This relationship has been obtained based on a simple approach that consists in the recursive application of invariance conditions, instead of relying on complicated concepts pertaining to group theory. This makes this approach versatile and accessible to a large audience. Moreover, to facilitate the application of the proposed approach, we have implemented a Python script that conveniently computes the form of the material parameter tensors and the scattering matrix directly from a list of specified symmetries. This Python script is freely accessible on GitHub~\cite{kagit}.

Based on this approach, we have shown how easy it is to investigate responses such as extrinsic chirality or asymmetric angular transmittance. We have also demonstrated the possibility of multipolar extrinsic chirality where chiral responses may be achieved in achiral structures even for normally incident waves due to the excitation of multipolar components; this was shown to result from the sensitivity of certain structures to field gradients. These examples demonstrate that intuition alone is insufficient to predict the effective material tensors and rich scattering behaviour that can be achieved, and thus highlight the usefulness of our method.

\section*{Acknowledgements}

We gratefully acknowledge funding from the Swiss National Science Foundation (project PZ00P2\_193221).

\appendix
\section{Symmetry Operations}
\label{appendix}

Reflection symmetries may be expressed using the Householder formula~\cite{householder1964}
\begin{equation}
	\label{eq_HM}
	\te{\Lambda} = \te{I} - 2\ve{n}\ve{n},
\end{equation}
where $\ve{n}$ corresponds to the reflection axis. It follows that the symmetry operations $\sigma_x$, $\sigma_y$ and $\sigma_z$ are defined by
\begin{subequations}
	\label{eq_pxpypz}
	\begin{align}
		\sigma_x &= \begin{bmatrix}
			-1 & 0 & 0 \\
			0 & 1 &  0 \\
			0 & 0  &  1 \\
		\end{bmatrix},\label{eq_px}\\
		\sigma_y &= \begin{bmatrix}
			1 & 0 & 0 \\
			0 & -1 &  0 \\
			0 & 0  &  1 \\
		\end{bmatrix},\\
		\sigma_z &= \begin{bmatrix}
			1 & 0 & 0 \\
			0 & 1 &  0 \\
			0 & 0  &  -1 \\
		\end{bmatrix}.
	\end{align}
\end{subequations}
On the other hand, the 3D rotation matrices are defined by
\begin{subequations}
	\begin{align}
		R_x(\theta) &= \begin{bmatrix}
			1 & 0 & 0 \\
			0 & \cos\theta &  -\sin\theta \\
			0 & \sin\theta  &  \cos\theta \\
		\end{bmatrix},\\
		R_y(\theta) &= \begin{bmatrix}
			\cos\theta & 0 & \sin\theta \\
			0 & 0 &  0 \\
			-\sin\theta & 0  &  \cos\theta \\
		\end{bmatrix},\\
		R_z(\theta) &= \begin{bmatrix}
			\cos\theta & -\sin\theta & 0 \\
			\sin\theta & \cos\theta &  0 \\
			0 & 0  &  1 \\
		\end{bmatrix}.\label{eq_rz}
	\end{align}
\end{subequations}
These matrices may be used to define the symmetry operation $C_{N,i}$ as
\begin{equation}
	\label{eq_Cni}
	C_{N,i} = R_i\left(\frac{2\pi}{N}\right).
\end{equation}

\bibliography{mybib}
	
\end{document}